\documentclass[a4paper]{article}

\usepackage{cmbright}
\usepackage{amsmath}
\usepackage{natbib}

\newtheorem{lemma}{Lemma}
\newtheorem{theorem}{Theorem}

\begin{document}

\title{Missing at random: a stochastic process perspective}

\author{Daniel Farewell \and Rhian Daniel \and Shaun Seaman}

\maketitle

\begin{abstract}
We offer a natural and extensible measure-theoretic treatment of
missingness at random. Within the standard missing data framework, we
give a novel characterisation of the observed data as a stopping-set
sigma algebra. We demonstrate that the usual missingness at random
conditions are equivalent to requiring particular stochastic processes
to be adapted to a set-indexed filtration of the complete data:
measurability conditions that suffice to ensure the likelihood
factorisation necessary for ignorability. Our rigorous statement of the
missing at random conditions also clarifies a common confusion: what is
fixed, and what is random?
\end{abstract}

Keywords: ignorability; missingness at random; sigma algebra; stochastic process.

\section{Introduction}\label{introduction}

Missing at random \citep{rubin_inference_1976} is a central concept in
missing data research. Nevertheless, recent papers
\citep{seaman_what_2013, mealli_clarifying_2015, doretti_missing_2017}
have argued that it remains poorly understood and often inaccurately
articulated. The most common formulation \citep[
p.12]{little_statistical_2002} is superficially intuitive but misleading
in its details; accurate formulations exist
\citep{robins_non-response_1997} but typically hold little heuristic
appeal.

Our ambition here is to provide both rigour and intuition. We show that
the factorization required for ignorability depends on the
data-measurability of the likelihood ratio between two probability
measures. The data themselves can be understood as a stopping-set sigma
algebra arising from a particular filtration of the probability space.
We give an explicit characterisation of two families of probability
measures: the first describes the measure assumed to be operating in
practice, while the second is a conditional version of the first under a
particular working independence assumption. We shall show that their
likelihood ratio is itself a stochastic process evaluated at a stopping
set, measurable if the underlying process is adapted. This leads
directly to a rigorous definition of missingness at random, but one that
retains the familiarity and simple appeal of the usual formulation.
Along the way, we draw out some deeper connections between missing data
problems and causal inference.

Rigour is relative. The level of formality we adopt is chosen for
clarity in matters we believe to be most important or least well
understood. For example, we express conditioning statements in terms of
sigma algebras, principally to avoid any confusion over precisely what
information is being conditioned upon. The fact that not only discrete
but also continuous or more general random variables can then be
subsumed within our setup is convenient, but secondary.

We tread this path with some trepidation. Rubin describes his own
initial measure-theoretic treatment of missing data as ``window
dressing'', and then-\emph{Biometrika}-editor David Cox's advice to him
was to ``eliminate all that measure theory noise''
\citep{rubin_past_2014}. We hope that our stochastic process perspective
avoids these pitfalls, and instead succeeds in exposing a real signal
that could to be overlooked following such noise reduction.

\section{Notation}\label{notation}

Our starting point is a measurable space \((\Omega, {\cal F})\) on which
we define various probability measures and random variables. We denote
probability measures by \(P\) and \(Q\), possibly indexed by a parameter
\(\theta\) in order to describe families of such measures. We assume
that all such probability measures are dominated by a known reference
measure \(\nu\): that is, if \(\nu(A) = 0\) for some set
\(A \in {\cal F}\), then also \(P(A) = Q(A) = 0\). Following
\citet{pollard_users_2002}, we adopt de Finetti notation: where the
context allows it, we interpret a set \(A\) as the random variable
\(1_{A}\), and we re-use the symbol \(P\) for its corresponding
expectation operator \(E_{P}\). Thus
\(P(A) = P(1_{A}) = E_{P}(1_{A}) = \int 1_{A}\; \mathrm{d}P\).

The sigma algebra \({\cal F}\) represents complete information about the
entire stochastic system. We think of the information provided by the
observed data, too, as a sigma algebra, and denote it by \({\cal D}\).
This sigma algebra \({\cal D}\) should contain all those events whose
logical status is known once the realised values of the observed data
are known. For the time being, we shall remain nebulous about the
precise definition of \({\cal D}\) but, given the missing data setting,
we shall expect \({\cal D}\) to be a strict subset of \({\cal F}\).

The usual definition of the likelihood function is in terms of the
Radon-Nikodym derivative \({{\rm d}P_{\theta}} / {{\rm d}\nu}\): the
density of \(P_{\theta}\) with respect to a dominating reference measure
\(\nu\). This density is a random variable that, for fixed
\(\omega \in \Omega\), is to be understood as a function of \(\theta\).
However, this random variable is not, in general,
\({\cal D}\)-measurable: in other words,
\({{\rm d}P_{\theta}} / {{\rm d}\nu}\) is the likelihood based on all
information in \({\cal F}\), and may depend on events whose truth or
fiction cannot be determined from the observed data \({\cal D}\). Such a
quantity is sometimes referred to as the complete data likelihood. By
contrast, the observed data likelihood may be conveniently represented
by \(\nu({{\rm d}P_{\theta}} / {{\rm d}\nu} \mid {\cal D})\): given
\({\cal D}\), the conditional expectation with respect to \(\nu\) of the
complete data likelihood \citep{commenges_likelihood_2005}. This has the
intuitive appeal of a local average of the complete data density over
the area of the sample space consistent with the observed data. As noted
by \citet[p.299]{chang_conditioning_1997}, it is also more economical
than the standard notation \(\int f\; \mathrm{d}y_{\mathrm{mis}}\),
since no \(y_{\mathrm{mis}}\) need be introduced.

\section{Ignorability}\label{ignorability}

Missingness at random is fundamentally concerned with ignorability: when
can two families of probability measures be used interchangeably for
likelihood-based inference about \(\theta\)? We shall make the natural
definition and assert that the families \((P_{\theta})\) and
\((Q_{\theta})\) are everywhere equivalent for inference about
\(\theta\) if, for all \(\theta, \theta'\),
\[\frac{\nu\left(\frac{{\rm d}P_{\theta}}{{\rm d}\nu} \mid {\cal
D}\right)}{\nu\left(\frac{{\rm d}P_{\theta'}}{{\rm d}\nu} \mid {\cal D}\right)}
= \frac{\nu\left(\frac{{\rm d}Q_{\theta}}{{\rm d}\nu} \mid {\cal
D}\right)}{\nu\left(\frac{{\rm d}Q_{\theta'}}{{\rm d}\nu} \mid {\cal
D}\right)}\] so that the two likelihood ratios are identical. Here and
elsewhere, equality should be understood almost surely. More simply,
writing \(f\) and \(g\) respectively for the observed-data likelihood
functions based on \(P\) and \(Q\), we are making the obvious assertion
that \(P\) and \(Q\) are equivalent for inference about \(\theta\) if it
is always the case that
\[\frac{f(\theta)}{f(\theta')} = \frac{g(\theta)}{g(\theta')}.\] It may
happen that everywhere equivalence does not hold, but that the two
families are equivalent on a subset \(A \in {\cal D}\). Formally,
\((P_{\theta})\) and \((Q_{\theta})\) are equivalent on
\(A \in {\cal D}\) for inference about \(\theta\) if, for all
\(\theta, \theta'\), \[\frac{A\nu\left(\frac{{\rm d}P_{\theta}}{{\rm
d}\nu} \mid {\cal D}\right)}{A\nu\left(\frac{{\rm d}P_{\theta'}}{{\rm d}\nu}
\mid {\cal D}\right)} = \frac{A\nu\left(\frac{{\rm d}Q_{\theta}}{{\rm d}\nu}
\mid {\cal D}\right)}{A\nu\left(\frac{{\rm d}Q_{\theta'}}{{\rm d}\nu} \mid
{\cal D}\right)};\] recall that \(A\) is to be interpreted here as
\(1_{A}\). This equality ensures that, for any \(\omega \in A\), the
realized values of likelihood ratios computed under \(P_{\theta}\) and
\(Q_{\theta}\) are identical. We now establish a simple condition under
which families of probability measures are equivalent in this sense.

\begin{lemma} The families $(P_{\theta})$ and $(Q_{\theta})$ are equivalent on
$A \in {\cal D}$ for inference about $\theta$ if $A \times {{\rm d}P_{\theta}}
/ {{\rm d}Q_{\theta}}$ is ${\cal D}$-measurable and does not vary with
$\theta$. The families are everywhere equivalent if ${{\rm d}P_{\theta}}
/ {{\rm d}Q_{\theta}}$ is ${\cal D}$-measurable and does not vary with
$\theta$. \end{lemma}

The proof is fairly direct: decomposing
\({{\rm d}P_{\theta}} / {{\rm d}\nu}\) as
\({{\rm d}P_{\theta}} / {{\rm d}Q_{\theta}} \times {{\rm d}Q_{\theta}} / {{\rm d}\nu}\),
\(A \in {\cal D}\) is brought inside the conditional expectations of the
definition and then
\(A \times {{\rm d}P_{\theta}} / {{\rm d}Q_{\theta}}\) taken outside
since, by assumption, it is \({\cal D}\)-measurable. More explicitly,
\[\frac{A\nu\left(\frac{{\rm d}P_{\theta}}{{\rm d}\nu} \mid {\cal
D}\right)}{A\nu\left(\frac{{\rm d}P_{\theta'}}{{\rm d}\nu} \mid {\cal
D}\right)} = \frac{\nu\left(A\frac{{\rm d}P_{\theta}}{{\rm
d}Q_{\theta}}\frac{{\rm d}Q_{\theta}}{{\rm d}\nu} \mid {\cal
D}\right)}{\nu\left(A\frac{{\rm d}P_{\theta'}}{{\rm d}Q_{\theta'}}\frac{{\rm
d}Q_{\theta'}}{{\rm d}\nu} \mid {\cal D}\right)} = \frac{A\frac{{\rm
d}P_{\theta}}{{\rm d}Q_{\theta}}\nu\left(\frac{{\rm d}Q_{\theta}}{{\rm d}\nu}
\mid {\cal D}\right)}{A\frac{{\rm d}P_{\theta'}}{{\rm
d}Q_{\theta'}}\nu\left(\frac{{\rm d}Q_{\theta'}}{{\rm d}\nu} \mid {\cal
D}\right)} = \frac{A\nu\left(\frac{{\rm d}Q_{\theta}}{{\rm d}\nu} \mid {\cal
D}\right)}{A\nu\left(\frac{{\rm d}Q_{\theta'}}{{\rm d}\nu} \mid {\cal
D}\right)},\] where the final cancellation of the two Radon-Nikodym
derivatives follows from the assumption that these do not vary with
\(\theta\). Taking \(A = \Omega\), a trivial corollary is that
\((P_{\theta})\) and \((Q_{\theta})\) are everywhere equivalent if
\({{\rm d}P_{\theta}} / {{\rm d}Q_{\theta}}\) is \({\cal D}\)-measurable
and does not vary with \(\theta\).

Reducing ignorability to a question about data-measurability of a
likelihood ratio such as \({{\rm d}P_{\theta}} / {{\rm d}Q_{\theta}}\)
is a very general idea. Indeed, to this point, we have made no mention
of missing data, and in fact the theory applies equally well in settings
where the observed data arise in a random fashion but unobserved
quantities are not thought of as data but simply as latent variables:
random effects, for example. This is the perspective taken by
\citet{farewell_ignorability_2017}.

Henceforth we shall suppress dependence on \(\theta\), and consider
conditions under which the measures \(P\) and \(Q\) implicitly defined
by \citet{rubin_inference_1976} satisfy this condition for inferential
equivalence.

\section{Monotone Missing Data}\label{monotone-missing-data}

\subsection{Data}\label{data}

Throughout the remainder of the paper, we employ the machinery and
methods of stochastic processes. For general missing data, the theory of
stochastic processes indexed by sets will be required
\citep{molchanov_theory_2006}. Here we begin with the gentler case of
monotone missingness, where it suffices to use standard theory for
stochastic processes in discrete time. Unlike other approaches, the
stochastic process perspective permits ideas to be extended from the
monotone case to the general setting with essentially trivial, semantic
modifications. Following \citet{rubin_inference_1976}, we let
\(Y = (Y_{1}, \ldots, Y_{n})\) be random variables defined on
\((\Omega, {\cal F})\), the ranges of which may be any measurable
spaces. We observe \(Y_{1}, \ldots, Y_{M}\), where the random variable
\(M\) satisfies \(0 \leq M \leq n\); we do not observe
\(Y_{M + 1}, \ldots, Y_{n}\). Observation of the stochastic process
\(Y\) is terminated at the random time \(M\).

A filtration is a nested family of sigma algebras that, heuristically,
captures the idea of information increase over time. The process \(Y\)
is adapted to its natural filtration \(({\cal Y}_{m})\), where the sigma
algebra \({\cal Y}_{m} = \sigma(Y_{i}: i \leq m)\). Similarly, \(M\) is
a stopping time with respect to the filtration \(({\cal M}_{m})\), where
\({\cal M}_{m} = \sigma(\{M \leq i\}: i \leq m)\). We define a larger
filtration \(({\cal F}_{m})\) by
\({\cal F}_{m} = {\cal Y}_{m} \vee {\cal M}_{m}\); this \({\cal F}_{m}\)
is the smallest sigma-algebra containing \({\cal Y}_{m}\) and
\({\cal M}_{m}\). For simplicity, we shall assume that
\({\cal F} = {\cal F}_{n}\), so that there are no measurable events
beyond those described by \(Y\) and \(M\). Similarly, we write
\({\cal Y} = {\cal Y}_{n}\) and \({\cal M} = {\cal M}_{n}\) to describe
complete information about \(Y\), or \(M\), respectively.

Loosely, \({\cal F}_{m}\) tells us the values of
\(Y_{1}, \ldots, Y_{m}\) and, through knowledge of the indicators
\(\{M \leq 1\}, \ldots, \{M \leq m\}\), whether or not we have stopped
recording measurements before time \(m\). The observed information
increases, \({\cal F}_{1} \subseteq \ldots \subseteq {\cal F}_{M}\),
until the random time \(M\), at which point no further information is
recorded. This idea is captured through the elegant definition of the
stopping time sigma algebra
\({\cal F}_{M} = \{A \in {\cal F}: A \cap \{M \leq m\} \in {\cal F}_{m} \mbox{ for all } m\}\).
This \({\cal F}_{M}\) is precisely what we mean by the observed data
sigma algebra \({\cal D}\). We contrast our formulation of
\({\cal D} = {\cal F}_{M}\) with the alternative
\({\cal D} = \sigma(M, Y_{1} \times \{M \geq 1\}, \ldots, Y_{n} \times \{M \geq n\})\)
which, while technically correct in certain circumstances, is less
revealing about the nature of the observed data. More importantly,
though, interpreting \({\cal D}\) as a stopping-time sigma algebra
provides us with weak and natural conditions for data-measurability.

\subsection{Probability Measures}\label{probability-measures}

We write \(P\) for the probability measure that we believe gives rise to
the data. As \citet{seaman_what_2013} point out, \(P\) need not actually
be the true data generating measure; \(P\) and \(Q\) are models, and
ignorability asks whether we can substitute the simpler model \(Q\) for
the more complex model \(P\) while obtaining identical inference about
\(\theta\). We assume that those aspects of \(P\) concerned with
specifying the marginal distribution of \(M\) are of little scientific
interest and, were \(Y\) and \(M\) independent, we would happily base
our inference on the conditional likelihood of \(Y\) given \(M\), since
this really only involves modelling the marginal distribution of \(Y\).

However, we concede that, under \(P\), the random variables \(Y\) and
\(M\) may in fact have a complicated dependence. Under what conditions
can we make a working assumption of independence between \(Y\) and
\(M\), use the associated conditional likelihood, and still draw the
same inferences as we would were we to use the full likelihood given by
\(P\)?

Specifically, let \(Q\) be any probability measure that dominates \(P\),
agrees with \(P\) on \({\cal Y}\), but under which \(Y\) and \(M\) are
in fact independent. That is, we assume that \(P(A) = Q(A)\) for all
\(A \in {\cal Y}\), and that \(Q(A \cap B) = Q(A)Q(B)\) for any
\(A \in {\cal Y}, B \in {\cal M}\). We shall assume that at least one
such \(Q\) exists; this depends essentially on \((\Omega, {\cal F})\)
having a suitable product structure. Denote by
\(Q': {\cal F} \times \Omega \to [0, 1]\) a regular conditional
probability given \({\cal M}\) that satisfies
\(Q'(A, \omega) = Q(A \mid {\cal M})(\omega)\). Our central question is
this: under what conditions are the likelihoods based on \(P\) and
\(Q'\) equivalent for inference about \(\theta\)?

Somewhat less explicitly, this is the question asked by
\citet{rubin_inference_1976}. We remark that our use of \(Q'\) to
characterize a conditional likelihood under a working independence
assumption is slightly different to that of
\citet{rubin_inference_1976}. Rubin's approach is to define the working
likelihood through a random probability measure that varies with \(M\),
under which the observed pattern of missing data is guaranteed to occur.
We feel that our definition is more charitable to the missing data
community: a working independence assumption seems more defensible than
a pretense that the observed pattern of missingness was the only
possibility. Another advantage is that our theory can more easily be
extended to cases where the range of \(M\) is uncountable; this is not
the case for the standard presentation
\citep[p.14]{commenges_likelihood_2005}. In practice, though, the two
approaches are equivalent.

We can be more specific about the relationship between \(P\) and \(Q'\).
The Radon-Nikodym derivative \({{\rm d}P} / {{\rm d}Q}\) is given by
\(\lambda / \mu\), where \(\lambda\) and \(\mu\) are conditional
densities of \(M\) given \({\cal Y}\) under \(P\) and \(Q\),
respectively. This is an emphatically causal notion
\citep{pearl_causality_2009}: we replace the conditional distribution of
\(M\) given \({\cal Y}\) under \(P\) with an alternative that does not
depend on \({\cal Y}\), calling the resulting measure \(Q\). The
relationship between \(Q\) and \(Q'\) is simpler still:
\({{\rm d}Q} / {{\rm d}Q'} = \mu\), because under \(Q\) the marginal
density of \(M\) and the conditional density of \(M\) given \({\cal Y}\)
are both \(\mu\). Consequently,
\({{\rm d}P} / {{\rm d}Q'} = {{\rm d}P} / {{\rm d}Q} \times {{\rm d}Q} / {{\rm d}Q'} = \lambda\).
We deduce that \(P\) and \(Q'\) are equivalent for inference if
\(\lambda\) is data-measurable and does not vary with \(\theta\).

\subsection{Measurability}\label{measurability}

We have characterized \(\lambda\), the likelihood ratio between \(P\)
and \(Q'\), as the conditional density of \(M\) given \({\cal Y}\): that
is, \(\lambda = \lambda_{M}\), where for each \(m\),
\(\lambda_{m} = P(M = m \mid {\cal Y})\). Looked at another way,
\(\lambda\) is the value taken by the stochastic process
\((\lambda_{m})\) at the random time \(M\).

This provides a direct route to the question of data measurability or,
more explicitly, \({\cal F}_{M}\)-measurability. For an
\(({\cal F}_{m})\)-stopping time \(M\), standard stochastic process
theory asserts that \(\lambda_{M}\) is \({\cal F}_{M}\)-measurable if
the process \((\lambda_{m})\) is adapted to the filtration
\(({\cal F}_{m})\): adaptedness means that each \(\lambda_{m}\) is
\({\cal F}_{m}\)-measurable. But since
\(\lambda_{m} = P(M = m \mid {\cal Y})\) must necessarily be
\({\cal Y}\)-measurable, for it to additionally be
\({\cal F}_{m}\)-measurable it must in fact be measurable with respect
to \(({\cal Y} \wedge {\cal F}_{m})\), the largest sigma algebra
contained in both. But \({\cal Y} \wedge {\cal F}_{m} = {\cal Y}_{m}\),
so in turn we must have \[P(M = m \mid
{\cal Y}) = P(M = m \mid {\cal Y}_{m})\mbox{ for every $m$.}\] This is
equivalent to the everywhere version of missingness at random
\citep{seaman_what_2013}, about which we make several comments. First,
its appearance is familiar. It looks strikingly similar to the
ubiquitous, informal definition
\(P(M \mid Y) = P(M \mid Y_{\mathrm{obs}})\)
\citep{little_statistical_2002}, yet its interpretation is rather
different: the condition applies to a sequence of fixed values \(m\),
not the random variable \(M\). Second, its appearance is simple,
particularly when contrasted with the rigorous definition given by
\citet{seaman_what_2013}; by conditioning on sub-sigma algebras, we
automatically demand that the equality hold for all possible data
consistent with the observed subset \(Y_{1}, \ldots, Y_{m}\). Third, it
applies equally well to a random variable \(Y\) taking values in
uncountable spaces: no conditioning on a set of measure zero is
required. Weaker versions of this condition are possible, which we now
briefly discuss.

\subsection{Realised MAR}\label{realised-mar}

It is useful to distinguish between everywhere and realised versions of
missingness at random since, for Bayesian or direct-likelihood
inference, only the realised likelihood function is relevant
\citep{doretti_missing_2017}. We consider the question of realised
ignorability with reference to the largest set \(A \in {\cal D}\) for
which \(A \lambda\) is \({\cal D}\)-measurable, which we may define as
the union of all sets satisfying this measurability condition.

Although it may have measure zero, this set is never empty: certainly
\(\{M = n\} \subseteq A\), since the set \(\{M = n\}\) is
\({\cal F}_{M}\)-measurable and the stochastic process
\((\{m = n\} \lambda_{m})\) is adapted to \(({\cal F}_{m})\), since
\(\lambda_{n} = P(M = n \mid {\cal Y}) = P(M = n \mid {\cal Y}_{n})\) is
a tautology. If \(\omega\) happens to fall in such a set \(A\), then
likelihood inference can proceed equivalently based on \(Q'\) instead of
\(P\).

\section{Non-monotone missing data}\label{non-monotone-missing-data}

We turn now to the general case, where there need be no natural ordering
of the \(Y_{i}\): the observations may be obtained simultaneously or in
an arbitrary order that is unknown to the observer. The variables
themselves may be of different types: some binary, some continuous, some
multivariate, and any possible subset of the variables may be observed.
Despite this generality, remarkably few notational changes are needed
from the ordered, monotone case: we simply reinterpret what we have
written to this point in terms of stochastic processes indexed by sets
\citep[p.334]{molchanov_theory_2006}. Our subscript \(i\) becomes a set,
so that if \(i = \{1, 3, 4\}\) then \(Y_{i} = (Y_{1}, Y_{3}, Y_{4})\).
The most important change from the monotone case is this: we now
understand \(M\) as a random subset of \(\{1, \ldots, n\}\),
representing the subset of variables that are observed.

There is, of course, no total ordering of the subsets of
\(\{1, \ldots, n\}\), but we exploit the partial ordering given by set
inclusion. That is, we interpret \(i \leq m\) as \(i \subseteq m\),
which describes a lattice on which the stochastic process \(Y\) is
defined. Once again, we stop observing \(Y\) at the random point \(M\)
on this lattice, but now there are potentially multiple routes by which
we may arrive at a given point. Just as before, however, we observe the
values of all random variables \(Y_{i}\) for which \(i \leq m\).

We define \({\cal Y}_{m} = \sigma(Y_{i}: i \leq m)\),
\({\cal M}_{m} = \sigma(\{M \leq i\}: i \leq m)\) and
\({\cal F}_{m} = {\cal Y}_{m} \vee {\cal M}_{m}\) just as before, where
now \(({\cal F}_{m})\) is a set-indexed filtration. Again as before,
\({\cal D} = {\cal F}_{M}\), a stopping set sigma algebra. The
probability measures \(P\), \(Q\) and \(Q'\) are unaltered in their
definitions, and the question of ignorability remains one of
\({\cal F}_{M}\)-measurability of \(\lambda = \lambda_{M}\), where
\(\lambda_{m} = P(M = m \mid {\cal Y})\). The everywhere missing at
random condition, too, is unaltered, and forms our central theorem,
which we now state formally.

\begin{theorem} Suppose $P(M = m \mid {\cal Y}) = P(M = m \mid {\cal Y}_{m})$
for all $m$. Let $Q \gg P$ agree with $P$ on ${\cal Y}$ and, under $Q$, let
${\cal Y}$ and ${\cal M}$ be independent. Define $Q'$ from $Q$ as a regular
conditional probability given ${\cal M}$, and let $\lambda_{m} = P(M = m \mid
{\cal Y})$. Then ${{\rm d}P} / {{\rm d}Q'} = \lambda = \lambda_{M}$ is ${\cal
F}_{M}$-measurable. \end{theorem}

The proof is identical to the monotone case: adaptedness of the process
\((\lambda_{m})\) to the filtration \(({\cal F}_{m})\) and the fact that
\(M\) is an \(({\cal F}_{m})\)-stopping time ensures that
\(\lambda = \lambda_{M}\) is \({\cal F}_{M}\)-measurable, as required.
Even in this general and unordered setting, the techniques and intuition
of stochastic processes provide us with a direct proof of
data-measurability of the likelihood ratio \({{\rm d}P} / {{\rm d}Q'}\).
The crucial point is adaptedness of the stochastic process
\((\lambda_{m})\): ignorability hangs on this natural condition.

\section{Discussion}\label{discussion}

At least initially, our aim in writing this paper was simply to provide
a rigorous re-interpretation of the usual missingness at random
formulation \(P(M \mid Y) = P(M \mid Y_{\mathrm{obs}})\) for those who,
like ourselves, worry about such things. We hope that our version,
\(P(M = m \mid {\cal Y}) = P(M = m \mid {\cal Y}_{m})\), fits this bill
and makes clear two things. First, there is absolutely no information
about \(M\) contained in either \({\cal Y}\) or \({\cal Y}_{m}\).
Second, the statement varies with different values of \(m\), and so is
not a global statement about the distribution of \(M\).

Missingness at random is thus certainly not a conditional independence
requirement \citep{mealli_clarifying_2015}. To our knowledge, our
characterization of missingness at random as a measurability requirement
and, in particular, as an adaptedness requirement, is novel. The
adaptedness condition is sufficient and nearly necessary: for a stopping
time \(M\), a random variable \(\lambda\) is \({\cal F}_{M}\)-measurable
if and only if there exists an adapted stochastic process \((X_{m})\)
such that \(\lambda = X_{M}\), so minimally \(\lambda_{M} = X_{M}\) for
some adapted process \((X_{m})\). We hope that the adaptedness
requirement will seem natural to those familiar with the theory of
stochastic processes, within which stochastic observation is very well
developed, and with which too few connections with missing data have
been made. For instance, the celebrated partial likelihood of survival
analysis \citep{cox_partial_1975} may be derived instead as a
conditional likelihood under a working independence assumption between
censoring, timings of failures and those individuals selected to fail.
Under weak assumptions about censoring, inference about the regression
parameters under the partial likelihood and the full model would be
ignorably different but for the fact that these parameters are shared
between both the timing and selection components of the model. This is a
good example of a case where
\({{\rm d}P_{\theta}} / {{\rm d}Q_{\theta}}\) does in fact vary with
\(\theta\).

Standard terminology speaks of missing at random `mechanisms'. This is
distinctly causal language of which we approve, but the conditional
density of \(M\) given \({\cal Y}\) does not necessarily have a natural
causal interpretation: in longitudinal settings, it seems highly
unlikely that future observations, however interpreted, should be
allowed to causally influence the occurrence of previous measurements,
even in principle. Conditional densities along more dynamic filtrations
such as
\(\{\emptyset, \Omega\} = {\cal F}_{0} \subseteq {\cal F}_{0} \vee {\cal Y}_{1} \subseteq \cdots \subseteq {\cal F}_{n - 1} \subseteq {\cal F}_{n - 1} \vee {\cal Y}_{n} \subseteq {\cal F}_{n} =  {\cal F}\)
could have more direct causal interpretation, and the relevant
likelihood ratio would be a product of Radon-Nikodym derivatives of
regular conditional probabilities, measurable if in fact the transitions
from \({\cal  F}_{m - 1} \vee {\cal Y}_{m}\) to \({\cal F}_{m}\)
depended only on \({\cal F}_{m - 1}\): sequential missingness at random.
Our suspicion is that if we started with a more appropriate causal
model, conditions for ignorability would be more easily assessed, and
more often deemed implausible.

More broadly, the key point remains one of data-measurability of a
likelihood ratio: a very general idea that goes beyond ideas of missing
data to encompass partially observed, coarsened or entirely latent
stochastic processes \citep{commenges_likelihood_2005}. Defining
filtrations on these richer spaces designed to have a causal
interpretation seems to us a promising approach: among other things, it
allows for the possibility that the thing to be measured can exist
without a measurement taking place \citep{farewell_ignorability_2017}.
Each \(\omega \in \Omega\) corresponds to a realisation of an entire
suite of random variables; even for a single individual, there is no
particular reason that their observed \(Y_{1}\) in a realisation in
which \(M = \{1\}\) should be the same \(Y_{1}\) as that from a
realisation where \(M = \{1, \ldots, n\}\).

We make use of random sets to extend the theory of monotone missing data
to general patterns of missingness. The lattice structure implicit in
this formulation is strongly reminiscent of the randomized monotone
missingness mechanisms of \citet{robins_non-response_1997}, wherein
future observation can depend on previous measurements within the
history of a particular branch. We believe, but have not proved, that
the distinction between randomized monotone missingness and general
missingness at random lies in the existence of incomparable sets
\(m, m'\), for which neither \(m \leq m'\) nor \(m' \leq m\).

\section*{Acknowledgement}
Odd Aalen, Vern Farewell and Robin Henderson provided valuable advice
during the preparation of this manuscript.

\bibliographystyle{biometrika}
\bibliography{onMAR}

\end{document}